\documentclass[11pt]{article}
\usepackage[english]{babel}
\usepackage[latin1]{inputenc}
\usepackage{amsmath,latexsym,amssymb}
\usepackage{amsfonts}
\usepackage{amsthm}
\usepackage{slashed}
\usepackage{color}


\newtheorem{thm}{Theorem}[section]
\newtheorem*{thm 1}{Theorem 1}
\newtheorem*{thm 2}{Theorem 2}
\newtheorem{cor}[thm]{Corollary}
\newtheorem{defi}[thm]{Definition}

\newtheorem{lem}[thm]{Lemma}
\newtheorem{prop}[thm]{Proposition}

\begin{document}

\title{Superfield equations in the Berezin-Kostant-Leites category}
\author{Michel Egeileh\footnotemark[1] \hspace{2mm} and \hspace{1mm} Daniel Bennequin\footnotemark[2]}
\date{\footnotemark[1] Conservatoire national des arts et m\'etiers (ISSAE Cnam Liban)\\
P.O. Box 113 6175 Hamra, 1103 2100 Beirut, Lebanon  \\  E-mail: {\it michel.egeileh@isae.edu.lb}\\
\vspace{3mm} \footnotemark[2] Institut de Math\'ematiques de Jussieu-Paris Rive Gauche,\\
 B\^atiment Sophie Germain, 8 place Aur\'elie Nemours, 75013 Paris, France  \\  E-mail: {\it bennequin@math.univ-paris-diderot.fr}\\}

\maketitle

\begin{abstract}
Using the functor of points, we prove that the Wess-Zumino equations for massive chiral superfields in dimension $4|4$ can be represented by supersymmetric equations in terms of superfunctions in the Berezin-Kostant-Leites sense (involving ordinary fields, with real and complex valued components). Then, after introducing an appropriate supersymmetric extension of the Fourier transform, we prove explicitly that these supersymmetric equations provide a realization of the irreducible unitary representations with positive mass and zero superspin of the super Poincar\'e group in dimension $4|4$.\\
\end{abstract}

\section{Introduction}
From the point of view of quantum field theory, $1$-particle states of a free elementary particle constitute a Hilbert space which, by the requirement of relativistic invariance, must carry an irreducible unitary representation of the Poincar\'e group $V\rtimes\hbox{Spin}(V)$, where $V$ is a Lorentzian vector space. In signature $(1,3)$, it is well-known since \cite{wig} that the irreducible representations of the Poincar\'e group that are of physical interest are classified by a nonnegative real number $m$ (the mass), and a half-integer $s\in\{0,\frac{1}{2},1,\frac{3}{2},...\}$ (the spin)\footnote[1] {the corresponding antiparticle representations are obtained by reversing the sign of the energy}. \\

It is also known, cf. \cite{bk}, that all the irreducible unitary representations of the Poincar\'e group of positive (resp. negative) energy and finite spectral multiplicity\footnote[2] {i.e. the representations that are induced from a finite-dimensional irreducible representation of the stabilizer of a point in an orbit of $\hbox{Spin}(V)$ in $V^{*}$
inside the closure of the positive (resp. negative) cone} are realizable as completions $\mathcal{H}$ of a quotient of a Hermitian linear subspace $\mathcal{F}$ of smooth sections of the vector bundle associated to a tensor-spinor representation
$W$ of $\hbox{Spin}(V)$ over the Minkowski spacetime $\mathbb{R}^d$ (i.e. the affine space directed by $V$). The subspace $\mathcal{F}$ is made of solutions
of special partial differential equations. For instance, if the orbit is the positive branch of hyperboloid defined by $|p|^{2}=m^{2}$ and the representation
$W$ is the space of Dirac spinors, the equation is the Dirac equation $\slashed D\psi+im\psi=0$. For the scalar representation, we get the
Klein-Gordon equation $d^{*}d\phi+m^{2}\phi=0$. If the orbit is the positive lightcone and $W$ is the complex vector representation, we find the Maxwell equation with the Lorentz condition, i.e. $d^{*}dA=0$ and $d^{*}A=0$. If we ask for real vectors, then two irreducible representations are needed, due to the symmetry of the Fourier transform: one of positive energy and one of negative energy. In this text, we will consider complex representations, but in subsequent works, it will be important to take into account the reality conditions on spinors and tensors which intervene in supersymmmetry. We will describe the construction of $\mathcal{H}$ with more details in sections $2$ and $5$.\\

The same kind of results holds for the irreducible representations of the super Poincar\'e group in the context of superfields, cf. \cite{wb}, \cite{bk}, \cite{dm}. In order to realize these representations in a space of superfields, one can consider {\it a priori} a suitable action functional for superfields on Minkowski superspacetime, and then obtain differential equations selecting the representation as the Euler-Lagrange equations corresponding to that action functional. This Lagrangian field-theoretic approach involves the differential geometry of the underlying supermanifold (here Minkowski superspacetime), in order to carry out the calculus of variations, and is most conveniently dealt with by applying the functor of points. This is what is implicitly done in the physics literature, and it leads naturally to spinor fields with anticommuting Grassmannian-valued components. For instance, one can obtain in this way the Wess-Zumino equations for massive chiral superfields (cf. \cite{wb}). \\

Starting with the category $\mathbf{sMan}$ of supermanifolds in the sense of Berezin, Kostant, Leites (cf. \cite{dm}), the extension corresponding to the functor of
points is given by the category $\mathbf{sF}$ of contravariant functors from $\mathbf{sMan}$ to the category $\mathbf{Set}$ of ordinary sets. There is an
embedding of $\mathbf{sMan}$ in $\mathbf{sF}$, given by associating to every supermanifold $M$ the representable functor $B\mapsto \hbox{Hom}_{\mathbf{sMan}}(B,M)$.
The necessity of this category for working with superfields and superfunctionals is apparent in the fact that functional spaces exist in $\mathbf{sF}$ but are
not representable in $\mathbf{sMan}$; for instance the simplest scalar superfields $\Phi$ from $M$ to $\mathbb{R}$ give rise to the functor which associates to
$B$ the set $\hbox{Hom}_{\mathbf{sMan}}(B\times M,\mathbb{R})$ (``inner Hom functor"), but no supermanifold corresponds to this functor.\\

Thus, it is \emph{a priori} a necessity to view the solutions of the equations associated to a representation of the super-Poincar\'e group as a functor.
However, the first result of this paper is that this functor is representable by the solutions of supersymmetric equations that involve only ordinary fields, with real or complex valued components. More precisely, with definitions and notations that will be given in Section 3, we have the following theorem:\\

\begin{thm 1} \label{representability}
Let $\mathcal{E}$ be the solution functor of the Wess-Zumino equations for massive chiral superfields, that is, the contravariant functor from the category of supermanifolds to the category of sets defined by

$\mathcal{E}(B):=\{\Phi:M_{cs}(B)\longrightarrow \mathcal{O}_B(|B|)_{\bar{0}}\;|\;\overline{D}_{\dot{a}}\Phi=0\;\hbox{ and }\;(D)^2\Phi-2m\,\overline{\Phi}=0\}$

 Then $\mathcal{E}$ is representable by the super-vector space

 $E:=\{f\in \mathcal{O}_{M_{cs}}(\mathbb{R}^4)\;|\;\overline{D}_{\dot{a}}f=0\;\hbox{ and }\;(D)^2 f-2m\,\overline{f}=0\}$

 in the sense that there is a natural isomorphism of functors: $\mathcal{E}\simeq\mathcal{L}E$,\\
where $\mathcal{L}E(B):=(E_0\hat{\otimes}\mathcal{O}_B(|B|)_{\bar{0}})\oplus (E_1\hat{\otimes}\mathcal{O}_B(|B|)_{\bar{1}})$ for every supermanifold $B$. Moreover, this isomorphism is super Poincar\'e-equivariant.\\
\end{thm 1}

The above theorem is a result about the nature of the fields. It is expected that the super-vector space $E$ should correspond, in a natural way, to an irreducible unitary representation of the super Poincar\'e group. Our second result will prove that this is indeed the case, i.e. the space $E$ of solutions of the above supersymmetric equations realizes the irreducible representation of super Poincar\'e with mass $m$ and superspin 0. More precisely, and again with definitions and notations that will be given in Section 3, we have the following theorem:\\

\begin{thm 2} \label{super equations}
Let $M_{cs}$ be the linear supermanifold associated to the super vector space $V\oplus S_{\mathbb{C}}$ (where $V$ is a four-dimensional Lorentzian vector space, and $S_{\mathbb{C}}$ the corresponding four-dimensional complex space of Dirac spinors). The irreducible unitary representation of the super-Poincar\'e algebra of mass $m$ and superspin 0 can be realized as the sub-super vector space of $\mathcal{O}_{M_{cs}}(\mathbb{R}^4)=\mathcal{C}^{\infty}(\mathbb{R}^4,\bigwedge^{\bullet}S_{\mathbb{C}}^*)$ made of the superfunctions satisfying the differential equations:\\

   $\overline{D}_{\dot{1}}f=0\;\;$, $\;\;\overline{D}_{\dot{2}}f=0\;\;$ and $\;\;(D)^2f=2m\,\bar{f}$.\\

In components, this representation space corresponds to:

 $\{(\varphi,\psi)\in \mathcal{C}^{\infty}(\mathbb{R}^4,\mathbb{C})\times \mathcal{C}^{\infty}(\mathbb{R}^4,S_+^*)\;|\;\left\{ \begin{array}{rcl}
(\square+m^2)\varphi &=&0 \\
\varepsilon^{ac}\Gamma_{a\dot{b}}^{\mu}\partial_{\mu}\psi_c-m\bar{\psi}_{\dot{b}} &=& 0 \\
\end{array} \right.\}\,.$\\
\end{thm 2}

The category $\mathbf{sF}$, containing functional spaces of superfields, provides an elegant formalism allowing to rewrite the functional super-calculus in
physics, which is particularly useful in quantum field theory, for instance in the computation of Batalin-Vilkoviski cohomology for quantization of gauge theories;
thus, it is legitimate to question the interest of an interpretation of supersymmetric particles in an ordinary geometrical manner.

A first reason comes from the interpretation of comparisons of the theory with possible experiments:
when the operator methods of QFT are applied, renormalized quantum amplitudes have to be expressed with ordinary
numbers, even if the operators belong to the super category. Then, the first interest of the geometric correspondence
is to make this fact more understandable.

A second reason, from a theoretical point of view, has to do with the interpretation of local interactions. First, is it possible to get ordinary nonlinear PDEs that are equivalent, in the sense of Theorems $1$ and $2$, to super-solutions of a supersymmetric super-Lagrangian? And, developing further the quantum theory of this supersymmetric super-Lagrangian, is it possible to get again a correspondence between the obtained quantum corrections and quantum corrections of an ordinary field theory? A negative answer would be particularly interesting.

Last but not least, the equations that appear, the linear ones and their nonlinear natural deformations, give new systems of partial differential equations which could be useful in differential geometry.\\

The first idea in our approach to proving Theorem 2 is a supersymmetric generalization of a construction that produces the Poincar\'e-invariant differential operators which cut out a given irreducible representation $\mathcal{H}$ in the space of spin-tensor fields. Roughly speaking, this construction can be described in the ordinary case as follows. Considering for instance the massive case, denote by $\mathcal{O}_m^+\subset V^*$ the orbit for $\hbox{Spin}(V)$ given by $|p|^2=m^2$ and lying in the forward timelike cone.
Then choose a preferred point in $\mathcal{O}_m^+$, and let $K$ be the stabilizer of that point. 
The representation $W$ of $\hbox{Spin}(V)$ that defines the spin-tensor fields can be restricted to $K$, which allows the construction of a $\hbox{Spin}(V)$-equivariant vector bundle over $\mathcal{O}_m^+$. We show in Section 3 that this bundle has a natural equivariant trivialization, which can be used to associate to every $K$-equivariant linear map from $W$ to another $\hbox{Spin}(V)$-module $E$, a $\hbox{Spin}(V)$-equivariant symbol $\zeta:\mathcal{O}_m^+\longrightarrow W^*\otimes E$. These symbols, in turn, give rise to Poincar\'e-equivariant differential operators on spacetime, by inverse Fourier transform.\\

The second idea is a suitable supersymmetric generalization of the Fourier transform, which we introduce and study in Section 6. One observation there is that the Hodge star can be interpreted as a purely odd super Fourier transform. \\

The article is organized as follows. In Section 2, we recall briefly the main result about the classification of irreducible unitary representations of the Poincar\'e group, which also serves as an opportunity to introduce our notations and terminology for the rest of the paper. In Section 3, we discuss the necessary notions to formulate Theorems 1 and 2. In Section 4, we prove Theorem 1. In Section 5, we explain in more detail our general construction of the Poincar\'e-invariant differential operators corresponding to a given irreducible representation. In Section 6, we introduce the relevant notion of super Fourier transform and discuss its main properties. In Section 7, we prove Theorem 2. Finally, Section 8 is an appendix in which we give a non-supersymmetric example (massive particle of arbitrary spin) that illustrates the construction in Section 5. \\

\section{Wigner representations}

In all this paper, $V$ denotes a real vector space of dimension $d$ equipped
with an inner product $\langle\,,\rangle$ of signature $(1,d-1)$. We choose one of the two connected components of the timelike cone $C=\{v\in V\;|\;\langle v,v\rangle>0\}$, and denote it by $C_+$, the other component being denoted by $C_-$. Dually, we set $C^{\vee}:=\{p\in V^*\;|\;\langle p,p\rangle>0\}$ and $C^{\vee}_{\pm}:=\{p\in C^{\vee}\;|\; p(v)\geq 0\;\;\forall v\in C_{\pm}\}$. Also, we denote by $\hbox{Spin}(V)$ the double cover of the connected Lorentz group of $V$ (preserving space and time orientation), and by $\Pi(V)$ the {\it Poincar\'e group} of $V$, defined as the semidirect product $V\rtimes\hbox{Spin}(V)$. \\

It is well-known that irreducible unitary representations of the Poincar\'e group are in one-to-one correspondence with pairs $(\mathcal{O},\sigma)$, where $\mathcal{O}$ is an orbit for the contragredient action of $\hbox{Spin}(V)$ on $V^*$, and $\sigma:K\longrightarrow\hbox{U}(F)$ a finite-dimensional irreducible unitary representation of the stabilizer $K$ of a preferred momentum $q\in\mathcal{O}$. This result goes back to Wigner for the Poincar\'e group (cf. \cite{wig}), and can be seen as a special case of a theorem of Mackey concerning the irreducible unitary representations of more general semidirect products of groups (cf. \cite{mac}). \\

For each $m>0$, the sheet of hyperboloid $\;\mathcal{O}_m^{\pm}:=\{p\in V^*\;|\;\langle p,p\rangle =m^2\}\cap C^{\vee}_{\pm}\;$ is an orbit; the parameter $m$ here is called the mass. Also, the one-sided light-cone $\;\mathcal{O}_0^{\pm}:=\{p\in V^*\;|\;\langle p,p\rangle =0\}\cap C^{\vee}_{\pm}\;$ is an orbit: it corresponds to the massless case. Irreducible unitary representations of the Poincar\'e group corresponding to $\mathcal{O}_m^+$ or $\mathcal{O}_0^+$ are called {\it of positive energy}.\\

When the orbit is $\mathcal{O}_m^+$ for some $m>0$, one can choose $q=me^0$ as preferred momentum. In this case, the stabilizer $K$ is isomorphic to $\hbox{Spin}(d-1)$. When the orbit is $\mathcal{O}_0^+$, one can choose $q=e^0+e^{d-1}$ as preferred momentum. In this case, the stabilizer is isomorphic to $\mathbb{R}^{d-2}\rtimes \hbox{Spin}(d-2)$, but only the (finite-dimensional) irreducible unitary representations of the compact part $\hbox{Spin}(d-2)$ of the stabilizer will lead to physically meaningful representations of the Poincar\'e group $\Pi(V)$. Thus, we will call {\it little group} and denote by $K$ either the stabilizer $\hbox{Spin}(d-1)$ in the massive case, or the compact quotient $\hbox{Spin}(d-2)$ of the stabilizer in the massless case.\\

The irreducible unitary representations of positive energy of the Poincar\'e group that come from a (finite-dimensional) irreducible unitary representation of the little group will be called {\it Wigner representations}.\\

\section{Superfield equations and super Poincar\'e representations}

In this section, we prepare the ground for and then formulate each of our results. In what follows, we restrict ourselves to $d=4$ (so $V$ is a vector space with an inner product of signature $(1,3)$). We will deal with representations in various dimensions in a subsequent extended article.\\

Denote by $S_{\mathbb{C}}$ the irreducible complex representation of the Clifford algebra $\hbox{C}\ell(V)$. Then $S_{\mathbb{C}}^*=S_+^*\oplus S_-^*$, and we fix once for all a basis $\{\theta^a\}$ in $S_+^*$ and $\{\bar{\theta}^{\dot{b}}\}$ in $S_-^*$, so that $S_+^*\simeq\mathbb{C}^2$ and $S_-^*\simeq\overline{\mathbb{C}^2}$. {\it Complex superspacetime} is defined as the supermanifold $M_{cs}=(\mathbb{R}^4,\mathcal{O}_{M_{cs}})$,
where $\mathcal{O}_{M_{cs}}(U):=\mathcal{C}^{\infty}(U,\mathbb{C})\otimes \bigwedge S_{\mathbb{C}}^*$ for every open
set $U\subset\mathbb{R}^4$.\\ A superfunction on $M_{cs}$ can be written:

$f=\varphi+\psi_a\theta^a+\eta_{\dot{b}}\bar{\theta}^{\dot{b}}+F\theta^1\theta^2+G\bar{\theta}^{\dot{1}}\bar{\theta}^{\dot{2}}+
\Gamma^{\mu}_{a\dot{b}}A_{\mu}\theta^a\bar{\theta}^{\dot{b}}+\lambda_{\dot{b}}\theta^1\theta^2\bar{\theta}^{\dot{b}}
+\mu_{a}\bar{\theta}^{\dot{1}}\bar{\theta}^{\dot{2}}\theta^a+H\theta^1\theta^2\bar{\theta}^{\dot{1}}\bar{\theta}^{\dot{2}}$

where $\varphi\;,\psi_a,\;\eta_{\dot{b}},\;F,\;G,\;A_{\mu},\;\lambda_{\dot{b}},\;\mu_{a},\;H\in \mathcal{C}^{\infty}(U,\mathbb{C})$ (we have $\hbox{dim}\bigwedge^{\bullet}S_{\mathbb{C}}^*=16$ complex-valued functions).

The global coordinates $(x^{\mu},\theta^a,\bar{\theta}^{\dot{b}})$ induce a global frame on $M_{cs}$: $\displaystyle (\frac{\partial}{\partial x^{\mu}},\frac{\partial}{\partial \theta^a},\frac{\partial}{\partial \bar{\theta}^{\dot{b}}})$,

but this is not the natural frame to consider: in fact, $M_{cs}$ has a natural super Lie group structure (with super Lie algebra $V\oplus S_{\mathbb{C}}$), though not an abelian one (supertranslations do not commute, as $[S_{\mathbb{C}},S_{\mathbb{C}}]\neq 0$). This leads to consider a global moving frame made of left-invariant super vector fields:

$P_{\mu}=\displaystyle\frac{\partial}{\partial x^{\mu}}\;\;$, $\;\;D_a=\displaystyle\frac{\partial}{\partial \theta^a} -\Gamma_{a\dot{b}}^{\mu}\bar{\theta}^{\dot{b}}\displaystyle\frac{\partial}{\partial x^{\mu}}\;\;$ and $\;\;\overline{D}_{\dot{a}}=\displaystyle\frac{\partial}{\partial \bar{\theta}^{\dot{a}}} -\Gamma_{b\dot{a}}^{\mu}\theta^b\displaystyle\frac{\partial}{\partial x^{\mu}}$

and we have: $[P_{\mu},D_a]=[P_{\mu},\overline{D}_{\dot{a}}]=0$, $[D_a,D_b]=[\overline{D}_{\dot{a}},\overline{D}_{\dot{b}}]=0$ and $[D_a,\overline{D}_{\dot{b}}]=2\,\Gamma_{a\dot{b}}^{\mu}P_{\mu}$.\\

The generalized supermanifold of superfields is $\mathcal{F}:=\underline{\hbox{Hom}}(M_{cs},\mathbb{C})$.
It is by definition the contravariant functor from the category of (finite-dimensional) supermanifolds $\mathbf{sMan}$ to $\mathbf{Set}$ given by
$$\mathcal{F}(B)=\hbox{Hom}(B\times M_{cs},\mathbb{C})$$ for all supermanifolds $B$.
One would like to think of $\mathcal{F}$ as some kind of infinite-dimensional supermanifold $(\mathcal{|F|},\mathcal{O}_{\mathcal{F}})$. \\

If $\Phi_{geom}=(\check{\Phi},\Phi^{\sharp}):B\times M_{cs}\longrightarrow \mathbb{C}$ is a morphism, then $\Phi^{\sharp}:\mathcal{C}^{\infty}(\mathbb{C},\mathbb{C})\longrightarrow \mathcal{O}_B(|B|)\;\hat{\otimes}\;\mathcal{O}_{M_{cs}}(\mathbb{R}^4)$ is entirely determined by $\Phi^{\sharp}(\hbox{\upshape{Id}}_{\mathbb{C}})\in (\mathcal{O}_B(|B|)\;\hat{\otimes}\;\mathcal{O}_{M_{cs}}(\mathbb{R}^4))_{\bar{0}}$. Thus, the information about $\Phi_{geom}$ is fully contained in $\Phi^{\sharp}(\hbox{\upshape{Id}}_{\mathbb{C}})\in (\bigwedge^{\bullet}S_{\mathbb{C}}^*\otimes\mathcal{C}^{\infty}(\mathbb{R}^4,\mathbb{C})\;\hat{\otimes}\;\mathcal{O}_B(|B|))_{\bar{0}}$.

Using $x^{\mu},\theta^a,\bar{\theta}^{\dot{b}}$ as coordinates on $M_{cs}$, we may write, for any $f\in\mathcal{C}^{\infty}(\mathbb{C},\mathbb{C})$,
$$\Phi^{\sharp}(f)=\varphi^{\sharp}(f)+\theta^a(\psi_a)^{\sharp}(f)+\bar{\theta}^{\dot{b}}(\eta_{\dot{b}})^{\sharp}(f)+\theta^1\theta^2 F^{\sharp}(f)+\bar{\theta}^{\dot{1}}\bar{\theta}^{\dot{2}}G^{\sharp}(f)+\Gamma^{\mu}_{a\dot{b}}\theta^a\bar{\theta}^{\dot{b}}(A_{\mu})^{\sharp}(f)$$
$$+\theta^1\theta^2\bar{\theta}^{\dot{b}}(\lambda_{\dot{b}})^{\sharp}(f)+\bar{\theta}^{\dot{1}}\bar{\theta}^{\dot{2}}\theta^a(\mu_a)^{\sharp}(f)+\theta^1\theta^2\bar{\theta}^{\dot{1}}\bar{\theta}^{\dot{2}}H^{\sharp}(f)$$

where $\varphi^{\sharp}(f)\in \mathcal{C}^{\infty}(\mathbb{R}^4,\mathbb{C})\;\hat{\otimes}\;\mathcal{O}_B(|B|)_{\bar{0}}$, $\;\; (\psi_a)^{\sharp}(f)\in \mathcal{C}^{\infty}(\mathbb{R}^4,\mathbb{C})\;\hat{\otimes}\;\mathcal{O}_B(|B|)_{\bar{1}}$ $\;\;\hbox{(while }\theta^a(\psi_a)^{\sharp}(f)\in S_+^*\otimes\mathcal{C}^{\infty}(\mathbb{R}^4,\mathbb{C})\;\hat{\otimes}\;\mathcal{O}_B(|B|)_{\bar{1}}$), etc...\\

To the superfield $\Phi_{geom}$, we can associate a map $\Phi:M_{cs}(B)\longrightarrow\mathcal{O}_B(|B|)_{\bar{0}}$, so that
$$\Phi(y,\xi,\bar{\xi})=\varphi(y)+\xi^a\psi_a(y)+ \bar{\xi}^{\dot{b}}\eta_{\dot{b}}(y)+\xi^1\xi^2 F(y)+\bar{\xi}^{\dot{1}}\bar{\xi}^{\dot{2}}G(y)+\Gamma^{\mu}_{a\dot{b}}\xi^a\bar{\xi}^{\dot{b}}A_{\mu}(y)$$
$$+\xi^1\xi^2\bar{\xi}^{\dot{b}}\lambda_{\dot{b}}(y)+\bar{\xi}^{\dot{1}}\bar{\xi}^{\dot{2}}\xi^a\mu_a(y)+\xi^1\xi^2\bar{\xi}^{\dot{1}}\bar{\xi}^{\dot{2}}H(y)$$
where $\varphi(y)\in\mathcal{O}_B(|B|)_{\bar{0}}$, $\xi^a\in\mathcal{O}_B(|B|)_{\bar{1}}$ and $\psi_a(y)\in \mathcal{O}_B(|B|)_{\bar{1}}$ (so $\xi^a\psi_a(y)\in\mathcal{O}_B(|B|)_{\bar{0}}$), etc...\\

Let $\Phi:M_{cs}(B)\longrightarrow\mathcal{O}_B(|B|)_{\bar{0}}$ be a scalar superfield. We say that $\Phi$ is {\it chiral} (resp. {\it antichiral}) if $\overline{D}_{\dot{1}}\Phi=\overline{D}_{\dot{2}}\Phi=0$ (resp. $D_1\Phi=D_2\Phi=0$). For any chiral superfield, let
$$\mathcal{A}(\Phi)=\int_{M_{cs}(B)}\bar{\Phi}\Phi\;d^2\xi\;d^2\bar{\xi}\;d^4 y+\int_{M_{cs}^+(B)}\frac{1}{2}m\Phi^2\;d^2\xi\;d^4 y$$
The superfield equation corresponding to the above action functional is $$(D)^2\Phi-2m\,\overline{\Phi}=0$$

Let $\mathcal{E}(B):=\{\Phi:M_{cs}(B)\longrightarrow \mathcal{O}_B(|B|)_{\bar{0}}\;|\;\overline{D}_{\dot{a}}\Phi=0\;\hbox{ and }\;(D)^2\Phi-2m\,\overline{\Phi}=0\}$\\
$\simeq \{(\varphi,\psi)\in \hbox{\upshape{Map}}(\mathbb{R}^4(B),\mathcal{O}_B(|B|)_{\bar{0}})\times \hbox{\upshape{Map}}(\mathbb{R}^4(B),(\mathcal{O}_B(|B|)_{\bar{1}})^2)\;|\; \left\{ \begin{array}{rcl}
(\square+m^2)\varphi &=&0 \\
\varepsilon^{ac}\Gamma_{a\dot{b}}^{\mu}\partial_{\mu}\psi_c-m\bar{\psi}_{\dot{b}} &=& 0 \\
\end{array} \right.\}$\\

$\mathcal{E}$ is the solution functor of the superfield equation. It is a generalized supermanifold on which the super Poincar\'e group $\hbox{S}\Pi(V)$ acts. \\

We are ready to state the first result of this paper, whose proof will be given in the next section.\\

\begin{thm 1} \label{representability}
Let $\mathcal{E}$ be the solution functor of the Wess-Zumino equations for massive chiral superfields, that is, the contravariant functor from the category of supermanifolds to the category of sets defined by

$\mathcal{E}(B):=\{\Phi:M_{cs}(B)\longrightarrow \mathcal{O}_B(|B|)_{\bar{0}}\;|\;\overline{D}_{\dot{a}}\Phi=0\;\hbox{ and }\;(D)^2\Phi-2m\,\overline{\Phi}=0\}$

 Then $\mathcal{E}$ is representable by the super-vector space

 $E:=\{f\in \mathcal{O}_{M_{cs}}(\mathbb{R}^4)\;|\;\overline{D}_{\dot{a}}f=0\;\hbox{ and }\;(D)^2 f-2m\,\overline{f}=0\}$

 in the sense that there is a natural isomorphism of functors: $\mathcal{E}\simeq\mathcal{L}E$,\\
where $\mathcal{L}E(B):=(E_0\hat{\otimes}\mathcal{O}_B(|B|)_{\bar{0}})\oplus (E_1\hat{\otimes}\mathcal{O}_B(|B|)_{\bar{1}})$ for every supermanifold $B$. Moreover, this isomorphism is super Poincar\'e-equivariant.\\
\end{thm 1}

This theorem shows that it is possible to obtain supersymmetric field equations as partial differential equations in terms of ordinary real or complex valued fields, without hidden anticommuting variables.\\

In fact, the above supersymmetric equations correspond in a natural way to an irreducible representation of the super Poincar\'e group. This is not very surprising, but a careful statement and proof are not available in the literature, up to our knowledge. This motivates our second result, stated below after introducing the necessary notations and terminology.\\

Let $S$ be an irreducible real spinorial representation of $\hbox{Spin}(V)$. We know (cf. Deligne's ``Notes on Spinors" in \cite{dm} or Chapter 6 in \cite{var1}) that there exists a $\hbox{Spin}(V)$-equivariant symmetric bilinear map $\Gamma:S\times S\longrightarrow V$. With this, one can define a super Lie algebra extending the Poincar\'e Lie algebra, by setting $\mathfrak{s}\pi(V):=\mathfrak{spin}(V)\oplus V\oplus S$, with the nontrivial brackets defined as follows: $[\mathfrak{spin}(V),S]$ is given by the action of $\mathfrak{spin}(V)$ on $S$, whereas $[V,S]=0$ and $[s_1,s_2]=-2\,\Gamma(s_1,s_2)$ for all $s_1,s_2\in S$. The super Lie algebra $\mathfrak{s}\pi(V)$ is called {\it super Poincar\'e algebra}.\\

Let $\rho:\mathfrak{s}\pi(V)\longrightarrow \mathfrak{gl}(\mathcal{H})$ be an irreducible unitary representation of the super Poincar\'e algebra on a super Hilbert space $\mathcal{H}=\mathcal{H}_{\bar{0}}\oplus \mathcal{H}_{\bar{1}}$. One can show that similarly to the non-super case, this representation is associated to a pair $(\mathcal{O},\sigma)$, where: $\mathcal{O}$ is an orbit of $\hbox{Spin}(V)$ on $V^*$ (except that now, this orbit will necessarily lie in the closure of $C_+^{\vee}$: supersymmetry forces the positivity of the energy), and $\sigma:\mathfrak{k}\oplus S\longrightarrow\mathfrak{gl}(F)$ is a finite-dimensional irreducible unitary representation of $\mathfrak{k}\oplus S$, where $\mathfrak{k}$ is the Lie algebra of the little group $K$. Given a preferred momentum $q\in\mathcal{O}$, there is a natural inner product on $S$, defined by $\langle s_1,s_2\rangle_q\;:=\;\frac{1}{2}q([s_1,s_2])\;=\;-q(\Gamma(s_1,s_2))$. Now since $\rho$ is a morphism of super Lie algebras, we have $[\rho(s_1),\rho(s_2)]=\rho([s_1,s_2])$, and on $F$, we have $\rho([s_1,s_2])=iq([s_1,s_2])\,\hbox{Id}_F$. It follows that $\;\rho(s_1)\circ \rho(s_2)+\rho(s_2)\circ \rho(s_1)\;=\;2i\langle s_1,s_2\rangle_q\;$, which shows that $F$ carries a representation of the Clifford algebra $\hbox{C}\ell(S,\langle\,,\rangle_q)$.\\

If $F$ is irreducible (as a Clifford module), then $F\simeq \bigwedge^{\bullet}\mathbb{C}^2$. Thus, $F=F_{\bar{0}}\oplus F_{\bar{1}}$, where $\;F_{\bar{0}}\;\simeq\;\bigwedge^{0}\mathbb{C}^2\oplus \bigwedge^{2}\mathbb{C}^2\;=\;\mathbb{C}\oplus\mathbb{C}\;$ and $\;F_{\bar{1}}\;\simeq\;\bigwedge^{1}\mathbb{C}^2=\mathbb{C}^2\;$ (as $K$-modules). We will consider the irreducible unitary representation $\mathcal{H}$ of the super Poincar\'e algebra $\mathfrak{s}\pi(V)$ that corresponds to $\mathcal{O}_m^+$ for some $m>0$ (``massive case") and $F\simeq \bigwedge^{\bullet}\mathbb{C}^2$ (``superspin 0").\\

\begin{thm 2} \label{super equations}
Let $M_{cs}$ be the linear supermanifold associated to the super vector space $V\oplus S_{\mathbb{C}}$ (where $V$ is a four-dimensional Lorentzian vector space, and $S_{\mathbb{C}}$ the corresponding four-dimensional complex space of Dirac spinors). The irreducible unitary representation of the super-Poincar\'e algebra of mass $m$ and superspin 0 can be realized as the sub-super vector space of $\mathcal{O}_{M_{cs}}(\mathbb{R}^4)=\mathcal{C}^{\infty}(\mathbb{R}^4,\bigwedge^{\bullet}S_{\mathbb{C}}^*)$ made of the superfunctions satisfying the differential equations:\\

   $\overline{D}_{\dot{1}}f=0\;\;$, $\;\;\overline{D}_{\dot{2}}f=0\;\;$ and $\;\;(D)^2f=2m\,\bar{f}$.\\

In components, this representation space corresponds to:

 $\{(\varphi,\psi)\in \mathcal{C}^{\infty}(\mathbb{R}^4,\mathbb{C})\times \mathcal{C}^{\infty}(\mathbb{R}^4,S_+^*)\;|\;\left\{ \begin{array}{rcl}
(\square+m^2)\varphi &=&0 \\
\varepsilon^{ac}\Gamma_{a\dot{b}}^{\mu}\partial_{\mu}\psi_c-m\bar{\psi}_{\dot{b}} &=& 0 \\
\end{array} \right.\}\,.$\\
\end{thm 2}

The proof of this theorem will be given in Section 7. It relies on two components. One of them is a supersymmetric version of a general construction that produces Poincar\'e-invariant differential operators which cut out a given Wigner representation in the space of spin-tensor fields. We will explain this construction in Section 5. The second component is a suitable supersymmetric generalization of the Fourier transform, which we will define and study in Section 6.\\

\section{Proof of Theorem 1}

Given a finite-dimensional super-vector space $V=V_{\bar{0}}\oplus V_{\bar{1}}$, we denote by $\hbox{L}V$ the supermanifold naturally associated to $V$ (``linear supermanifold"), so that $\hbox{L}V=(V_{\bar{0}},\mathcal{O}_{LV})$, where $\mathcal{O}_{LV}(U):=\mathcal{C}^{\infty}(U)\otimes \bigwedge^{\bullet}V_1^*$ for every open set $U\subset V_{\bar{0}}$. \\

On the other hand, given a super-vector space $V=V_{\bar{0}}\oplus V_{\bar{1}}$ (not necessarily finite-dimensional), we can associate to $V$ a functor $\mathcal{L}V:\mathbf{sMan}^{op}\longrightarrow\mathbf{Set}$ (``generalized supermanifold"), defined by $\mathcal{L}V(B):=(\mathcal{O}_B(|B|)_{\bar{0}}\,\hat{\otimes}\, V_{\bar{0}})\oplus (\mathcal{O}_B(|B|)_{\bar{1}}\,\hat{\otimes}\, V_{\bar{1}})$ for every supermanifold $B$. Note that if $V$ is finite-dimensional, then this functor is representable in the category $\mathbf{sMan}$, precisely by the linear supermanifold $\hbox{L}V$: we have $\hbox{Hom}(\cdot, \hbox{L}V)\simeq \mathcal{L}V$ (natural isomorphism of functors).\\

Now let $M$ be a supermanifold and $W=W_{\bar{0}}\oplus W_{\bar{1}}$ a finite-dimensional super-vector space. To define supersymmetric linear sigma-models, one needs a notion of ``space of maps" from $M$ to the linear supermanifold $\hbox{L}W$. Such a space of maps should constitute some kind of infinite-dimensional supermanifold; in this article, we treat such spaces as functors (generalized supermanifolds). Thus, we consider the inner hom functor $\underline{\hbox{Hom}}(M,\hbox{L}W):\mathbf{sMan}^{op}\longrightarrow\mathbf{Set}$, defined by $\underline{\hbox{Hom}}(M,\hbox{L}W)(B):=\hbox{Hom}(B\times M,\hbox{L}W)$.\\

One naturally expects that in a suitable category of infinite-dimensional supermanifolds, $\underline{\hbox{Hom}}(M,\hbox{L}W)$ should be representable by an infinite-dimensional {\it linear} supermanifold. Since we are working with functors, this leads us to expect that there should be a natural isomorphism of functors $\underline{\hbox{Hom}}(M,\hbox{L}W)\simeq \mathcal{L}V$ for some infinite-dimensional vector space $V=V_{\bar{0}}\oplus V_{\bar{1}}$. The following lemma shows that this indeed the case.\\

\begin{lem} Let $M$ be a supermanifold and $W=W_{\bar{0}}\oplus W_{\bar{1}}$ a finite-dimensional super-vector space. We have the following natural isomorphism of functors $$\underline{\hbox{\upshape{Hom}}}(M,\hbox{\upshape{L}}W)\;\simeq\;\mathcal{L}(\mathcal{O}_M(|M|)\otimes W)$$
\end{lem}

\noindent {\bf Proof.~} For every supermanifold $B$,

 $\underline{\hbox{\upshape{Hom}}}(M,\hbox{\upshape{L}}W)(B)\;=\;\hbox{Hom}(B\times M,\hbox{L}W)\;\simeq\;\hbox{Hom}_{\mathbf{sAlg}}(\mathcal{O}_{LW}(W_{\bar{0}})\,,\,\mathcal{O}_{B}(|B|)\,\hat{\otimes}\,\mathcal{O}_M(|M|))$

 $\simeq\;\hbox{Hom}_{\mathbf{sVect}}(W^*\,,\,\mathcal{O}_{B}(|B|)\,\hat{\otimes}\,\mathcal{O}_M(|M|))\;\simeq\;(\mathcal{O}_{B}(|B|)\,\hat{\otimes}\,\mathcal{O}_M(|M|)\otimes W)_{\bar{0}}$

$\simeq\;(\mathcal{O}_{B}(|B|)_{\bar{0}}\,\hat{\otimes}\,(\mathcal{O}_M(|M|)\otimes W)_{\bar{0}})\,\oplus\,(\mathcal{O}_{B}(|B|)_{\bar{1}}\,\hat{\otimes}\,(\mathcal{O}_M(|M|)\otimes W)_{\bar{1}})$

$\simeq\;\mathcal{L}(\mathcal{O}_M(|M|)\otimes W)(B)$ $\hfill\square$\\

Now recall that Minkowski superspacetime is the linear supermanifold $M_{cs}=\hbox{L}(V\oplus S_{\mathbb{C}})$, thus $M_{cs}=(\mathbb{R}^4,\mathcal{O}_{M_{cs}})$ where we think of $\mathbb{R}^4$ as the linear manifold corresponding to $V$ and $\mathcal{O}_{M_{cs}}(U)=\mathcal{C}^{\infty}(U)\otimes\bigwedge^{\bullet}S_{\mathbb{C}}^*$ for every open set $U\subset \mathbb{R}^4$. Applying the previous lemma with $M=M_{cs}$ and $W=\mathbb{R}$ gives
$$\underline{\hbox{\upshape{Hom}}}(M_{cs},\mathbb{R})\;\simeq\;\mathcal{L}(\mathcal{O}_{M_{cs}}(\mathbb{R}^4))$$

Now if $E=E_{\bar{0}}\oplus E_{\bar{1}}$ is the sub-super-vector space of $\mathcal{O}_{M_{cs}}(\mathbb{R}^4)$ made of the solutions of the equations $\overline{D}_{\dot{a}}f=0\;\hbox{and}\;(D)^2 f-2m\,\overline{f}=0$, the functor $\mathcal{L}E$ is a subfunctor of $\mathcal{L}(\mathcal{O}_{M_{cs}}(\mathbb{R}^4))$. Also, the solution functor $\mathcal{E}$ of the Wess-Zumino equations for massive chiral superfields is a subfunctor of $\underline{\hbox{\upshape{Hom}}}(M_{cs},\mathbb{R})$. Now for every supermanifold $B$, the bijection $\mathcal{L}(\mathcal{O}_{M_{cs}}(\mathbb{R}^4))(B)\longrightarrow \underline{\hbox{\upshape{Hom}}}(M_{cs},\mathbb{R})(B)$ coming from the above natural isomorphism clearly sends $\mathcal{L}E(B)$ to $\mathcal{E}(B)$. Therefore, the above natural isomorphism induces a natural isomorphism between the subfunctors $\mathcal{L}E$ and $\mathcal{E}$, and it is not difficult to see that the latter preserves the action of the super Poincar\'e group.

\section{Field equations corresponding to Wigner representations}

We will describe in general terms how, for a given Wigner representation $\mathcal{H}$ of the Poincar\'e group, one can find a suitable space of classical fields, and a set of linear partial differential equations on those fields whose space of solutions corresponds to $\mathcal{H}$ in a natural way. \\

More precisely, consider a finite-dimensional representation of the group $\hbox{Spin}(V)$ on some complex vector space $W$ (i.e., a ``spin-tensor" representation). We may define a {\it classical field of type $W$} on Minkowski spacetime to be a compactly supported smooth section of the equivariant vector bundle $\mathbb{W}:=\Pi(V)\times_{Spin(V)}W$, associated by this representation to the principal $\hbox{Spin}(V)$-bundle $\Pi(V)\longrightarrow \mathbb{R}^d$. The space $\Gamma_c(\mathbb{R}^d,\mathbb{W})$ of classical fields of type $W$ carries a (highly reducible) representation of the Poincar\'e group $\Pi(V)$. Ultimately, one would be interested in finding some decomposition of this space of fields into irreducible representations. One step in this direction is to start with a Wigner representation $\mathcal{H}$, find a suitable spin-tensor representation $W$, and construct a natural inner product on $\Gamma_c(\mathbb{R}^d,\mathbb{W})$ in such a way that after separating (i.e. dividing out the kernel of that inner product) and completing, the resulting Hilbert space contains a subspace unitarily isomorphic to the Wigner representation $\mathcal{H}$ we started with.\\

The first question is the choice of the spin-tensor representation $W$. As will be clear from the following construction, one must choose the representation $W$ of $\hbox{Spin}(V)$ so that when restricted to the little group $K$, it contains a subrepresentation isomorphic to the irreducible unitary representation $\sigma:K\longrightarrow\hbox{U}(F)$ that defines $\mathcal{H}$. Once such a choice of $W$ has been made, if one's goal is to pick $\mathcal{H}$ out of $\Gamma_c(\mathbb{R}^d,\mathbb{W})$ (as representations of $\Pi(V)$), it is reasonable to try first to perform the ``algebraic analog", that is, pick $F$ out of $W$ (as representations of $K$). One way to achieve this is to have a $K$-equivariant linear map $u$ from $W$ to some other representation $E$ of $\hbox{Spin}(V)$, such that $\hbox{Ker}\,u\simeq F$. As for the inner product, one starts also by ``working at the algebraic level", by choosing an appropriate $K$-invariant Hermitian form $\langle\,,\rangle_0$ on $W$, whose restriction to $F$ is positive-definite. \\

Next, one extends the previous algebraic considerations, which can be thought of as taking place in the category of finite-dimensional representations of $K$, to an equivalent category of equivariant vector bundles. Although the inner product we have in mind can be ultimately defined on the space $\Gamma_c(\mathbb{R}^d,\mathbb{W})$ of all fields (or on their Fourier transforms, which is easier), we will content ourselves here with an on-shell description in momentum space.  Thus, for $m\geq 0$, we consider the equivariant vector bundle $\widehat{\mathbb{W}}_{(m)}:=\hbox{Spin}(V)\times_K W$ on $\mathcal{O}_m^+$, associated to the principal $K$-bundle $\hbox{Spin}(V)\longrightarrow \mathcal{O}_m^+$ by the restricted representation of $K$ on $W$. The $K$-equivariant linear map $u:W\longrightarrow E$ defines then a morphism of equivariant vector bundles $\widehat{\mathbb{W}}_{(m)}\longrightarrow \widehat{\mathbb{E}}_{(m)}$ (where $\widehat{\mathbb{E}}_{(m)}:=\hbox{Spin}(V)\times_K E$), which, in turn, defines a linear map between sections $\Gamma(\mathcal{O}_m^+,\widehat{\mathbb{W}}_{(m)})\longrightarrow  \Gamma(\mathcal{O}_m^+,\widehat{\mathbb{E}}_{(m)})$. The Wigner representation $\mathcal{H}$ corresponds then to the kernel of this linear map, and the inner product on $\mathcal{H}$ corresponds to the inner product on $\Gamma(\mathcal{O}_m^+,\widehat{\mathbb{W}}_{(m)})$ defined by the $K$-invariant Hermitian form on $W$. Finally, taking the inverse Fourier transform of the condition defining $\mathcal{H}$ in $\Gamma(\mathcal{O}_m^+,\widehat{\mathbb{W}}_{(m)})$ gives the desired partial differential equations.\\

In fact, one does not need to work with sections of vector bundles: there is a natural, equivariant trivialization of the the vector bundles $\widehat{\mathbb{W}}_{(m)}$ and $\widehat{\mathbb{E}}_{(m)}$. The reason is that these equivariant vector bundles are in fact ``tractor bundles", that is, they are associated to representations of $K$ which are in fact restrictions of representations of $\hbox{Spin}(V)$. In such a situation, there is always an equivariant trivialization: $\widehat{\mathbb{W}}_{(m)}\longrightarrow \mathcal{O}_m^+\times W$, given by $[h,w]\mapsto (hq,hw)$ (and, similarly, $\widehat{\mathbb{E}}_{(m)}\longrightarrow \mathcal{O}_m^+\times E$, given by $[h,e]\mapsto (hq,he)$). One can check that in these trivializations, the bundle morphism $\widehat{\mathbb{W}}_{(m)}\longrightarrow \widehat{\mathbb{E}}_{(m)}$ becomes the equivariant map $\zeta_u:\mathcal{O}_m^+\longrightarrow W^*\otimes E$ given by $\zeta_u(p)=h_p\cdot u\cdot h_p^{-1}$, where $h_p\in \hbox{Spin}(V)$ is such that $h_p\,q=p$. Then, $\mathcal{H}$ corresponds to the subspace of maps $\hat{\Phi}:\mathcal{O}_m^+\longrightarrow W$ satisfying the condition $\zeta_u(p)(\hat{\Phi}(p))=0$. \\

Typically, the spin-tensor representation $W$ is of real type, i.e. it admits a $\hbox{Spin}(V)$-invariant conjugation. Then $\mathcal{H}$ would also correspond to the space of maps $\hat{\Phi}:\mathcal{O}_m\longrightarrow W$ satisfying the reality condition $\hat{\Phi}(-p)=\overline{\hat{\Phi}(p)}$. By inverse Fourier transform, one gets fields valued in a real vector space, satisfying partial differential equations corresponding to the condition $\zeta_u(p)(\hat{\Phi}(p))=0$.\\

As for the inner product, one can check that it is given on the space of maps from $\mathcal{O}_m^+$ to $W$ by $\displaystyle\langle \hat{\Phi},\hat{\Psi}\rangle=\int_{\mathcal{O}_m^+} \langle h_p^{-1}\hat{\Phi}(p),h_p^{-1}\hat{\Psi}(p)\rangle_0\;d\beta_m^+(p)$, where $\beta_m^+$ is the $\hbox{Spin}(V)$-invariant volume form on $\mathcal{O}_m^+$ (and $h_p\in \hbox{Spin}(V)$ is such that $h_p\,q=p$).\\

This construction can be used to produce in a natural way the field equations corresponding to the usual irreducible representations of the Poincar\'e group, and so one can obtain in this way the Klein-Gordon equation, the Dirac equation and the Maxwell equation. More generally, we illustrate this construction by producing the field equations in dimension 4 that correspond to a massive particle of arbitrary spin (cf. the appendix, Section 8).\\

\section{Super Fourier transform}

Notions of Fourier transform in superspace have appeared previously in the literature, namely in \cite{dew}, \cite{rog} and \cite{deb}. These authors work in a different category of supermanifolds, and \cite{dew}, \cite{rog} use a kernel that transforms under the orthogonal group. On the other hand, \cite{deb} uses instead a kernel that transforms under the group $\hbox{Sp}(2n,\mathbb{R})$ (here, $n$ is the odd dimension). Inspired by this, we use
the standard supermetric on $M_{cs}$ to define a natural version of the Fourier
transform for Minkowski superspacetime, taking superfunctions in
$\mathcal{C}^{\infty}_c(\mathbb{R}^4)[\theta^a,\bar{\theta}^{\dot{a}}]$ to
superfunctions in $\mathcal{C}^{\infty}(V^*)[\tau^a,\bar{\tau}^{\dot{a}}]$. Once an
expression for the super Fourier transform of a superfunction is obtained, we
see that the purely odd part of the transform coincides with the Hodge
isomorphism defined by the invariant symplectic structure on the spinors. From
this, it is easy to check that the super Fourier transform has natural
properties such as exchanging the odd derivative
$\displaystyle\frac{\partial}{\partial\theta^1}$ with exterior multiplication
by $i\tau^2$, and multiplication by $\theta^1$ with the contraction
$\displaystyle -i\frac{\partial}{\partial\tau^2}$. The $1\leftrightarrow 2$ exchange is not surprising since the super Fourier transform is defined via a symplectic structure. Finally,
we define supersymmetric symbols $\zeta_{\bar{d}_{\bar{\tau}^{\dot{a}}}}$ and $\zeta_{d_{\tau^a}}$ that correspond, via the super Fourier transform, to the supertranslation-invariant odd vector fields $D_a$ and $\overline{D}_{\dot{a}}$ canonically defined on $M_{cs}$.\\

\begin{defi} The {\bf super Fourier transform} of a (compactly supported) superfunction\\ $f\in \mathcal{C}^{\infty}_c(\mathbb{R}^4,\bigwedge^{\bullet}S_{\mathbb{C}}^*)\simeq \mathcal{C}^{\infty}_c(\mathbb{R}^4)[\theta^a,\bar{\theta}^{\dot{a}}]$ is the element $\star\widehat{f}\in \mathcal{C}^{\infty}(V^*)[\tau^a,\bar{\tau}^{\dot{a}}]$ defined by:
$$\star\widehat{f}:=\int_{M_{cs}} e^{-i(\langle p,x\rangle+\varepsilon_+(\tau,\theta)+\varepsilon_-(\bar{\tau},\bar{\theta}))}\;f\;dx\;d\theta\;d\bar{\theta}$$
\end{defi}

Note that if we define the bosonic Fourier transform of $f$ by $\,\displaystyle\widehat{f}(p):=\int_{\mathbb{R}^4} e^{-i\langle p,x\rangle}\;f(x)\;dx\,$, then $\,\displaystyle\star\widehat{f}(p)=\int e^{-i(\varepsilon_+(\tau,\theta)+\varepsilon_-(\bar{\tau},\bar{\theta}))}\;\widehat{f}(p)\;d\theta\;d\bar{\theta}$.\\

\begin{prop} \label{super FT expression}
Let $\;\;f(x)\;\;=\;\;\varphi(x)\;+\;\psi_a(x)\,\theta^a\;+\;\eta_{\dot{b}}(x)\,\bar{\theta}^{\dot{b}}\;+\;F(x)\,\theta^1\wedge\theta^2\;+\;G(x)\,\bar{\theta}^{\dot{1}}\wedge\bar{\theta}^{\dot{2}}$

$+\;\Gamma_{a\dot{b}}^{\mu}\,A_{\mu}(x)\,\theta^a\otimes\bar{\theta}^{\dot{b}}\;+\;\lambda_{\dot{b}}(x)\,(\theta^{1}\wedge\theta^{2})\otimes\bar{\theta}^{\dot{b}}\;+\;\mu_a(x)\,\theta^a\otimes(\bar{\theta}^{\dot{1}}\wedge\bar{\theta}^{\dot{2}})\;+\;H(x)\,(\theta^1\wedge\theta^2)\otimes(\bar{\theta}^{\dot{1}}\wedge\bar{\theta}^{\dot{2}})$

be the expansion of a generic superfunction $f:\mathbb{R}^4\longrightarrow \bigwedge^{\bullet}S_{\mathbb{C}}^*$. Then the super Fourier transform of $f$ is given by\\

$\star\widehat{f}(p)\;\;=\;\;\widehat{H}(p)\;+\;i\widehat{\mu}_a(p)\,\tau^a\;+\;i\widehat{\lambda}_{\dot{b}}(p)\,\bar{\tau}^{\dot{b}}\;+\;\widehat{G}(p)\,\tau^1\wedge\tau^2\;+\;\widehat{F}(p)\,\bar{\tau}^{\dot{1}}\wedge\bar{\tau}^{\dot{2}}$

$-\;\Gamma_{a\dot{b}}^{\mu}\,\widehat{A}_{\mu}(p)\,\tau^a\otimes\bar{\tau}^{\dot{b}}\;+\;i\widehat{\eta}_{\dot{b}}(p)\,(\tau^{1}\wedge\tau^{2})\otimes\bar{\tau}^{\dot{b}}\;+\;i\widehat{\psi}_a(p)\,\tau^a\otimes(\bar{\tau}^{\dot{1}}\wedge\bar{\tau}^{\dot{2}})\;+\;\widehat{\varphi}(p)\,(\tau^1\wedge\tau^2)\otimes(\bar{\tau}^{\dot{1}}\wedge\bar{\tau}^{\dot{2}})$
\end{prop}

\noindent {\bf Proof.~}

We have $\;\;\widehat{f}(p)\;\;=\;\;\widehat{\varphi}(p)\;+\;\widehat{\psi}_a(p)\,\theta^a\;+\;\widehat{\eta}_{\dot{b}}(p)\,\bar{\theta}^{\dot{b}}\;+\;\widehat{F}(p)\,\theta^1\wedge\theta^2\;+\;\widehat{G}(p)\,\bar{\theta}^{\dot{1}}\wedge\bar{\theta}^{\dot{2}}$

$+\;\Gamma_{a\dot{b}}^{\mu}\,\widehat{A}_{\mu}(p)\,\theta^a\otimes\bar{\theta}^{\dot{b}}\;+\;\widehat{\lambda}_{\dot{b}}(p)\,(\theta^{1}\wedge\theta^{2})\otimes\bar{\theta}^{\dot{b}}\;+\;\widehat{\mu}_a(p)\,\theta^a\otimes(\bar{\theta}^{\dot{1}}\wedge\bar{\theta}^{\dot{2}})\;+\;\widehat{H}(p)\,(\theta^1\wedge\theta^2)\otimes(\bar{\theta}^{\dot{1}}\wedge\bar{\theta}^{\dot{2}})$

To derive the stated expression for the super Fourier transform $\star\widehat{f}(p)$, start by expanding the exponential $\;e^{-i(\varepsilon_+(\tau,\theta)+\varepsilon_-(\bar{\tau},\bar{\theta}))}=e^{-i\varepsilon_+(\tau,\theta)}e^{-i\varepsilon_-(\bar{\tau},\bar{\theta})}=e^{-i(\tau^1\theta^2-\tau^2\theta^1)}e^{-i(\bar{\tau}^{\dot{1}}\bar{\theta}^{\dot{2}}-\bar{\tau}^{\dot{2}}\bar{\theta}^{\dot{1}})}$

$=(1-i\tau^1\theta^2+i\tau^2\theta^1+\tau^1\tau^2\theta^1\theta^2)\;(1-i\bar{\tau}^{\dot{1}}\bar{\theta}^{\dot{2}}+i\bar{\tau}^{\dot{2}}\bar{\theta}^{\dot{1}}+\bar{\tau}^{\dot{1}}\bar{\tau}^{\dot{2}}\bar{\theta}^{\dot{1}}\bar{\theta}^{\dot{2}})\,$

Then, multiply the result by the expansion of $\widehat{f}(p)$, retaining only the coefficients of $(\theta^1\wedge\theta^2)\otimes(\bar{\theta}^{\dot{1}}\wedge\bar{\theta}^{\dot{2}})$, and finally perform a Berezin integration. $\hfill\square$\\

\begin{cor} \label{odd super Fourier equals Hodge star}
The purely odd super Fourier transform coincides with the Hodge dual (with respect to the symplectic form $\varepsilon$ on $S_{\mathbb{C}}$).
\end{cor}

\noindent {\bf Proof.~}
Just compare the expansions of $\widehat{f}(p)$ and $\star\widehat{f}(p)$. $\hfill\square$\\

\begin{prop}\
\begin{enumerate}
\item $\displaystyle \star(\widehat{\frac{\partial f}{\partial\theta^a}})=i\varepsilon_{ab}\,\tau^b\,(\star\widehat{f})\quad$,$\quad\displaystyle\star(\widehat{\frac{\partial f}{\partial\bar{\theta}^{\dot{a}}}})=i\varepsilon_{\dot{a}\dot{b}}\,\bar{\tau}^{\dot{b}}\,(\star\widehat{f})$
\item $\displaystyle \star(\widehat{\theta^af})=-i\varepsilon^{ab}\,\frac{\partial}{\partial\tau^b}\,(\star\widehat{f})\quad$,$\quad\displaystyle\star(\widehat{\bar{\theta}^{\dot{a}}f})=-i\varepsilon^{\dot{a}\dot{b}}\,\frac{\partial}{\partial\bar{\tau}^{\dot{b}}}\,(\star\widehat{f})$
\end{enumerate}
\end{prop}

\noindent {\bf Proof.~}
Follows from Proposition \ref{super FT expression}. $\hfill\square$\\

Recall that for every superfunction $f:\mathbb{R}^4\longrightarrow \bigwedge^{\bullet}S_{\mathbb{C}}^*$,

 we have $\;\;\displaystyle D_{a}f=\frac{\partial f}{\partial\theta^{a}}-\Gamma_{a\dot{b}}^{\mu}\bar{\theta}^{\dot{b}}\frac{\partial f}{\partial x^{\mu}}\;\;$ and $\;\;\displaystyle\overline{D}_{\dot{a}}f=\frac{\partial f}{\partial\bar{\theta}^{\dot{a}}}-\Gamma_{b\dot{a}}^{\mu}\theta^b\frac{\partial f}{\partial x^{\mu}}$.

 Also, we set $\;\;\displaystyle (D)^2f:=\frac{1}{2}\varepsilon^{ab}D_{a}D_{b}f\;\;$ and $\;\;\displaystyle (\overline{D})^2f:=\frac{1}{2}\varepsilon^{\dot{a}\dot{b}}\overline{D}_{\dot{a}}\overline{D}_{\dot{b}}f$.\\

We define the following symbols, acting on elements of $\mathcal{C}^{\infty}(V^*)[\tau^a,\bar{\tau}^{\dot{a}}]$:

$\zeta_{d_a}(p):=(i\varepsilon_{ab}\,e_{\tau^b}\,\otimes\,\hbox{Id})-(\hbox{Id}\,\otimes\,\varepsilon^{\dot{c}\dot{d}}\,\Gamma_{a\dot{c}}^{\mu}\,p_{\mu}\,\iota_{\bar{\tau}^{\dot{d}}})$

 $\zeta_{\bar{d}_{\dot{a}}}(p):=(\hbox{Id}\,\otimes\, i\varepsilon_{\dot{a}\dot{b}}\,e_{\bar{\tau}^{\dot{b}}})-(\varepsilon^{cd}\,\Gamma_{c\dot{a}}^{\mu}\,p_{\mu}\,\iota_{\tau^d}\,\otimes\,\hbox{Id})$

 $\displaystyle\zeta_{(d)^2}:=\frac{1}{2}\,\varepsilon^{ab}\,\zeta_{d_a}\circ\zeta_{d_b}\;\;$ and $\;\;\displaystyle\zeta_{(\bar{d})^2}:=\frac{1}{2}\,\varepsilon^{\dot{a}\dot{b}}\,\zeta_{\bar{d}_{\dot{a}}}\circ\zeta_{\bar{d}_{\dot{b}}}$.\\

\begin{prop}\
\begin{enumerate}
\item $\displaystyle\star\widehat{D_{a}f}(p)=\zeta_{d_a}(p)(\star\widehat{f}(p))\quad$,$\quad\displaystyle\star\widehat{\overline{D}_{\dot{a}}f}(p)=\zeta_{\bar{d}_{\dot{a}}}(p)(\star\widehat{f}(p))$
\item $\displaystyle\star\widehat{(D)^2f}(p)=\zeta_{(d)^2}(p)(\star\widehat{f}(p))\quad$,$\quad\displaystyle\star\widehat{(\overline{D})^2f}(p)=\zeta_{(\bar{d})^2}(p)(\star\widehat{f}(p))$
\end{enumerate}
\end{prop}

\noindent {\bf Proof.~}
Follows directly from the preceding proposition. $\hfill\square$\\

\section{Proof of Theorem 2}

The superfields we are considering are elements of the super-vector space $\mathcal{O}_{M_{cs}}(\mathbb{R}^4)=\mathcal{C}^{\infty}(\mathbb{R}^4)\otimes\bigwedge^{\bullet}S_{\mathbb{C}}^*=\mathcal{C}^{\infty}(\mathbb{R}^4,\bigwedge^{\bullet}S_{\mathbb{C}}^*)$, so adopting the notations of Section 3, we have here $W\;=\;\bigwedge^{\bullet}S_{\mathbb{C}}^*\;=\;\bigwedge^{\bullet}S_+^*\otimes\bigwedge^{\bullet}S_-^*\;$, since $\;S_{\mathbb{C}}=S_+\oplus S_-$.  \\

On the other hand, we have seen in Section 4 that the irreducible unitary representation $\mathcal{H}$ of the super-Poincar\'e algebra $\mathfrak{s}\pi(V)$ that is of mass $m>0$ and superspin 0 corresponds to the irreducible $\hbox{C}\ell(S,\langle\,,\rangle_q)$-module $F=\bigwedge^{\bullet}\mathbb{C}^2$.\\

As explained in Section 3, we need to a way to ``pick $F$ out of $W$" as representations of the stabilizer, and so we need some equivariant linear map on $W$ whose kernel would be $F$. Now $W$ is a Clifford module, but it is certainly reducible and a multiple of the irreducible Clifford module $F$. In fact, upon restriction to $K\simeq\hbox{Spin}(3)$, we get $S_+\simeq S_-$, and with a convenient choice of bases, one can write

$W\;=\;\bigwedge^{\bullet}\mathbb{C}^2\otimes\bigwedge^{\bullet}\mathbb{C}^2\;=\;\bigwedge^{\bullet}\mathbb{C}^2\otimes(\mathbb{C}\oplus\mathbb{C}^2\oplus\bigwedge^{2}\mathbb{C}^2)\;=\;\bigwedge^{\bullet}\mathbb{C}^2\oplus(\bigwedge^{\bullet}\mathbb{C}^2\otimes\mathbb{C}^2)\oplus\bigwedge^{\bullet}\mathbb{C}^2$

This direct sum of three terms shows that we have at least two ways of realizing $F$ inside $W$, one by choosing the $\bigwedge^{\bullet}\mathbb{C}^2$ occurring as the first term (``chiral choice"), and the other by choosing the $\bigwedge^{\bullet}\mathbb{C}^2$ occurring as the third term (``antichiral choice").\\

A canonical way to perform these choices is via the following linear maps. Recall that $\langle s_1,s_2\rangle_q=-q(\Gamma(s_1,s_2))$, so with $q=me^0$, we have $\langle s_1,s_2\rangle_q=-me^0(\Gamma(s_1,s_2))$. We set $\Gamma^0:=e^0\circ\Gamma$, and define the following endomorphisms of $W$:

$d_a:=(i\varepsilon_{ab}\,e_{\tau^b}\,\otimes\,\hbox{Id})-(\hbox{Id}\,\otimes\,\varepsilon^{\dot{c}\dot{d}}\,\Gamma_{a\dot{c}}^{0}\,\iota_{\bar{\tau}^{\dot{d}}})$

 $\bar{d}_{\dot{a}}:=(\hbox{Id}\,\otimes\, i\varepsilon_{\dot{a}\dot{b}}\,e_{\bar{\tau}^{\dot{b}}})-(\varepsilon^{cd}\,\Gamma_{c\dot{a}}^{0}\,\iota_{\tau^d}\,\otimes\,\hbox{Id})$

This allows to consider $F$ as the ``chiral subspace" $\hbox{Ker}\,\bar{d}_{\dot{1}}\cap\hbox{Ker}\,\bar{d}_{\dot{2}}$, which is in fact $K$-invariant. \\

Thus, we see that chirality of superfields in dimension $4|4$ appears already at the purely algebraic level. The next step is to proceed in the spirit of Section 3, associating actual symbols to the ``algebraic symbols" $d_a$ and  $\bar{d}_{\dot{a}}$. Not surprisingly, this gives exactly the symbols $\zeta_{d_a}(p)$ and $\zeta_{\bar{d}_{\dot{a}}}(p)$ defined in Section 5.\\

It remains to calculate the action of these symbols on the expansions of the superfields in momentum space, and impose the right mass-shell equation, involving the second-order symbol $\displaystyle\zeta_{(d)^2}(p)$.\\

\begin{lem} \label{chiral superfield}
A superfunction $\star\widehat{f}\in \mathcal{C}^{\infty}(V^*)[\tau^a,\bar{\tau}^{\dot{a}}]$ satisfies the condition $$\zeta_{\bar{d}_{\dot{a}}}(p)(\star\widehat{f}(p))=0$$
if and only if $\;\;\widehat{\mu}_a=\widehat{G}=\widehat{\eta}_{\dot{b}}=0\;$, $\;\widehat{H}(p)=\eta^{\mu\nu}\,ip_{\mu}\,\widehat{A}_{\nu}(p)\;$, $\;\widehat{A}_{\mu}(p)=-ip_{\mu}\,\widehat{\varphi}(p)$

and $\;\;\widehat{\lambda}_{\dot{b}}(p)=\varepsilon^{ac}\,\Gamma_{a\dot{b}}^{\mu}\,ip_{\mu}\,\widehat{\psi}_c(p)\;$, so that \\

$\star\widehat{f}(p)\;\;=\;\;|p|^2\,\widehat{\varphi}(p)\;-\;\varepsilon^{ac}\,\Gamma_{a\dot{b}}^{\mu}\,p_{\mu}\,\widehat{\psi}_c(p)\,\bar{\tau}^{\dot{b}}\;+\;\widehat{F}(p)\,\bar{\tau}^{\dot{1}}\wedge\bar{\tau}^{\dot{2}}\;+\;\Gamma_{a\dot{b}}^{\mu}\,ip_{\mu}\,\widehat{\varphi}(p)\,\tau^a\otimes\bar{\tau}^{\dot{b}}$

$+\;i\widehat{\psi}_a(p)\,\tau^a\otimes(\bar{\tau}^{\dot{1}}\wedge\bar{\tau}^{\dot{2}})\;+\;\widehat{\varphi}(p)\,(\tau^1\wedge\tau^2)\otimes(\bar{\tau}^{\dot{1}}\wedge\bar{\tau}^{\dot{2}})$
\end{lem}

\noindent {\bf Proof.~}
Applying $\;\displaystyle\zeta_{\bar{d}_{\dot{a}}}(p)=i\varepsilon_{\dot{a}\dot{b}}\,\tau^{\dot{b}}-\varepsilon^{cd}\,\Gamma_{c\dot{a}}^{\mu}\,p_{\mu}\,\frac{\partial}{\partial \tau^d}\;$ to the expansion of $\star\widehat{f}(p)$ given in Proposition \ref{super FT expression}, the result follows after a calculation, taking into account that we have $\Gamma_{1\dot{1}}^{\mu}\Gamma_{2\dot{2}}^{\nu}-\Gamma_{1\dot{2}}^{\mu}\Gamma_{2\dot{1}}^{\nu}=\eta^{\mu\nu}$. $\hfill\square$\\

\begin{lem}
Let $\star\widehat{f}\in \mathcal{C}^{\infty}(V^*)[\tau^a,\bar{\tau}^{\dot{a}}]$ be a superfunction satisfying the condition of the preceding lemma. Then\\

$\zeta_{(d)^2}(p)(\star\widehat{f}(p))\;\;=\;\;-|p|^2\,\widehat{F}(p)\;+\;(\varepsilon^{\dot{c}\dot{d}}\,\varepsilon^{dc}\,\Gamma_{a\dot{c}}^{\mu}\,\Gamma_{d\dot{d}}^{\nu}\,p_{\mu}\,p_{\nu}\,i\widehat{\psi}_c(p)-|p|^2\,i\widehat{\psi}_a(p))\,\tau^a$

$-\;4\,|p|^2\,\widehat{\varphi}(p)\,\tau^1\wedge\tau^2\;+\;\Gamma_{a\dot{b}}^{\mu}\,ip_{\mu}\,\widehat{F}(p)\,\tau^a\otimes\bar{\tau}^{\dot{b}}\;+\;2\,\Gamma_{a\dot{b}}^{\mu}\,p_{\mu}\,\widehat{\psi}_c(p)\,(\tau^a\wedge\tau^c)\otimes \bar{\tau}^{\dot{b}}$

$-\;\widehat{F}(p)\,(\tau^1\wedge\tau^2)\otimes(\bar{\tau}^{\dot{1}}\wedge\bar{\tau}^{\dot{2}})$
\end{lem}

\noindent {\bf Proof.~}
Applying $\;\displaystyle\zeta_{d_{a}}(p)=i\varepsilon_{ab}\,\tau^{b}-\varepsilon^{\dot{c}\dot{d}}\,\Gamma_{a\dot{c}}^{\mu}\,p_{\mu}\,\frac{\partial}{\partial \bar{\tau}^{\dot{d}}}\;$ to the expansion of $\star\widehat{f}(p)$ given in Lemma \ref{chiral superfield} gives

$\displaystyle\zeta_{d_{a}}(p)(\star\widehat{f}(p))\;\;=\;\;\varepsilon_{ab}\,|p|^2\,i\widehat{\varphi}(p)\,\tau^b\;-\;\varepsilon_{ab}\,\varepsilon^{dc}\,\Gamma_{d\dot{b}}^{\mu}\,ip_{\mu}\,\widehat{\psi}_c(p)\,\tau^b\otimes\bar{\tau}^{\dot{b}}$

$-\;\varepsilon_{ab}\,\Gamma_{c\dot{b}}^{\mu}\,p_{\mu}\,\widehat{\phi}(p)\,(\tau^b\wedge\tau^c)\otimes\bar{\tau}^{\dot{b}}\;+\;\varepsilon_{ab}\,i\widehat{F}(p)\,\tau^b\otimes(\bar{\tau}^{\dot{1}}\wedge\bar{\tau}^{\dot{2}})\;-\;\varepsilon_{ab}\,\widehat{\psi}_c(p)\,(\tau^b\wedge\tau^c)\otimes(\bar{\tau}^{\dot{1}}\wedge\bar{\tau}^{\dot{2}})$

$+\;\varepsilon^{\dot{c}\dot{d}}\,\varepsilon^{dc}\,\Gamma_{a\dot{c}}^{\mu}\,\Gamma_{d\dot{d}}^{\nu}\,p_{\mu}\,p_{\nu}\,\widehat{\psi}_c(p)\;+\;\varepsilon^{\dot{c}\dot{d}}\,\Gamma_{a\dot{c}}^{\mu}\,\Gamma_{c\dot{d}}^{\nu}\,p_{\mu}\,p_{\nu}\,i\widehat{\varphi}(p)\,\tau^c\;+\;\Gamma_{a\dot{c}}^{\mu}\,p_{\mu}\,\widehat{F}(p)\,\bar{\tau}^{\dot{c}}$

$-\;\Gamma_{a\dot{c}}^{\mu}\,ip_{\mu}\,\widehat{\psi}_c(p)\,\tau^c\otimes\bar{\tau}^{\dot{c}}\;+\;\Gamma_{a\dot{c}}^{\mu}\,p_{\mu}\,\widehat{\varphi}(p)\,(\tau^1\wedge\tau^2)\otimes\bar{\tau}^{\dot{c}}$

The next step is to apply to the above another instance of $\zeta_{d_{a}}(p)$, then contract with the symplectic form $\varepsilon$ on $S_+$ and divide by 2. After calculation and simplification, one obtains the stated expression for $\zeta_{(d)^2}(p)(\star\widehat{f}(p))$. $\hfill\square$\\

To conclude the proof of Theorem \ref{super equations}, we impose the equation
$$\zeta_{(d)^2}(p)(\star\widehat{f}(p))\;=\;2m\,\overline{\star\widehat{f}(p)}$$
Comparing the expansions of both sides, we see that

 $\;-\widehat{F}(p)=2m\,\overline{\widehat{\varphi}(p)}\;$ and $\;-4\,|p|^2\,\widehat{\varphi}(p)=2m\,\overline{\widehat{F}(p)}$

which gives $$(|p|^2-m^2)\,\widehat{\varphi}(p)\;=\;0$$
Since $\;2\,\Gamma_{a\dot{b}}^{\mu}\,p_{\mu}\,\widehat{\psi}_c(p)\,(\tau^a\wedge\tau^c)\otimes \bar{\tau}^{\dot{b}}\;=\;2\,\varepsilon^{ac}\,\Gamma_{a\dot{b}}^{\mu}\,p_{\mu}\,\widehat{\psi}_c(p)\,(\tau^1\wedge\tau^2)\otimes\bar{\tau}^{\dot{b}}\,$, we also obtain $$\varepsilon^{ac}\,\Gamma_{a\dot{b}}^{\mu}\,p_{\mu}\,\widehat{\psi}_c(p)\;=\;-im\,\overline{\widehat{\psi}_{\dot{b}}(p)}$$

Finally, taking the inverse super Fourier transform gives the equations in Theorem 2.\\

\section{Appendix: an example in dimension 4 with arbitrary spin} \label{examples}

In this appendix, we illustrate the content of Section 5 with an example in four-dimensional Minkowski spacetime: the case of a particle of mass $m>0$ and of spin $s\geq 1$.\\

We know that the corresponding Wigner representation $\mathcal{H}_{(m,s)}$ is obtained by induction from the irreducible unitary representation of $K\simeq\hbox{Spin}(3)\simeq\hbox{SU}(2)$ of spin $s$. The space of this representation is $F_{s}\simeq\hbox{Sym}^{2s}\mathbb{C}^2$.

We want to realize $\mathcal{H}_{(m,s)}$ in a space of spin-tensor fields $\mathcal{C}^{\infty}(\mathbb{R}^4,W)$. We can take $W=\hbox{Sym}^{2\alpha}S_+\otimes\hbox{Sym}^{2\beta}S_-$ with $\alpha+\beta\geq s$. We consider the minimal choice $\alpha+\beta=s$. Of course, $\alpha,\beta\in\{0,\frac{1}{2},1,\frac{3}{2},2,...\}$, and we consider in what follows the interesting case $\alpha>0$ and $\beta>0$. We need an $\hbox{SU}(2)$-equivariant map $u$ on the $\hbox{Spin}(V)$-module $W$ whose kernel is $F_{s}$. At this point, we need to know how to decompose $\hbox{Sym}^{2\alpha}S_+^*\otimes\hbox{Sym}^{2\beta}S_-^*$ into irreducible representations of $\hbox{SU}(2)$. \\

  Recall that the irreducible complex representations of $\hbox{SU}(2)$ are classified by $\{0,\frac{1}{2},1,\frac{3}{2},2,...\}$. More precisely, for each $s\in \{0,\frac{1}{2},1,\frac{3}{2},2,...\}$, the vector space $\hbox{Sym}^{2s}S_+^*$ carries the irreducible representation of $\hbox{SU}(2)$ of highest weight $s$. We say that $\hbox{Sym}^{2s}S_+^*$ is the irreducible representation of spin $s$ of $\hbox{SU}(2)$; its dimension is $2s+1$, and its internal structure can be described as follows.

A canonical choice of Cartan subalgebra of $\mathfrak{su}(2)$ is $\mathfrak{t}:=\{ \left( \begin{array}{cc} i\theta & 0 \\ 0 & -i\theta \end{array} \right)\;;\;\theta\in\mathbb{R} \}$. Under $\mathfrak{t}$, the representation $\hbox{Sym}^{2s}S_+^*$ decomposes into one-dimensional weight spaces:
 $$\hbox{Sym}^{2s}S_+^*=L_{-s}\oplus L_{-s+1}\oplus L_{-s+2}\oplus\dots\oplus L_{s-2}\oplus L_{s-1}\oplus L_{s}$$
where $\left( \begin{array}{cc} i\theta & 0 \\ 0 & -i\theta \end{array} \right)$ acts on $L_j$ by $\lambda\mapsto 2j(i\theta)\lambda$.\\

For instance, the spin 0 representation is the trivial representation on $\mathbb{C}$, the spin $\frac{1}{2}$ representation is $S_+^*=L_{-\frac{1}{2}}\oplus L_{\frac{1}{2}}$, and the spin 1 representation is $\hbox{Sym}^{2}S_+^*=L_{-1}\oplus L_0\oplus L_1$.\\

\begin{lem} \label{decomposition lemma}
Assume in addition that $\alpha\geq\beta$. Then, as representation of $\hbox{\upshape{SU}}(2)$,
$$\hbox{\upshape{Sym}}^{2\alpha}S_+^*\otimes\hbox{\upshape{Sym}}^{2\beta}S_-^*\;\simeq\; \hbox{\upshape{Sym}}^{2(\alpha+\beta)}S_+^*\;\oplus\; \hbox{\upshape{Sym}}^{2(\alpha+\beta-1)}S_+^* \;\oplus\; \dots \;\oplus\; \hbox{\upshape{Sym}}^{2(\alpha-\beta)}S_+^*$$
\end{lem}

\noindent {\bf Proof.~} As representations of $\hbox{SU}(2)$, $S_+^*$ and $S_-^*$ become equivalent. Thus, we need to decompose $\hbox{Sym}^{2\alpha}S_+^*\otimes\hbox{Sym}^{2\beta}S_+^*$. We start by writing the weight-space decomposition of each factor: we have
$$\hbox{Sym}^{2\alpha}S_+^*=L_{-\alpha}\oplus L_{-\alpha+1}\oplus L_{-\alpha+2}\oplus\dots\oplus L_{\alpha-2}\oplus L_{\alpha-1}\oplus L_{\alpha}$$
and
$$\hbox{Sym}^{2\beta}S_+^*=L_{-\beta}\oplus L_{-\beta+1}\oplus L_{-\beta+2}\oplus\dots\oplus L_{\beta-2}\oplus L_{\beta-1}\oplus L_{\beta}$$
Taking the tensor product, and using the fact that $L_j\otimes L_k=L_{j+k}$, we obtain
$$\hbox{Sym}^{2\alpha}S_+^*\otimes\hbox{Sym}^{2\beta}S_+^*=L_{-\alpha-\beta}\oplus 2L_{-\alpha-\beta+1}\oplus 3L_{-\alpha-\beta+2}\oplus\dots\oplus 3L_{\alpha+\beta-2}\oplus 2L_{\alpha+\beta-1}\oplus L_{\alpha+\beta}$$
which implies easily the result. $\hfill\square$\\

Notice that by the above lemma, $$\hbox{\upshape{Sym}}^{2\alpha-1}S_+^*\otimes\hbox{\upshape{Sym}}^{2\beta-1}S_-^*\;\simeq\; \hbox{\upshape{Sym}}^{2(\alpha+\beta-1)}S_+^*\;\oplus\; \hbox{\upshape{Sym}}^{2(\alpha+\beta-2)}S_+^* \;\oplus\; \dots \;\oplus\; \hbox{\upshape{Sym}}^{2(\alpha-\beta)}S_+^*$$
and therefore we have (also by the above lemma):
$$\hbox{\upshape{Sym}}^{2\alpha}S_+^*\otimes\hbox{\upshape{Sym}}^{2\beta}S_-^*\;\simeq\; \hbox{\upshape{Sym}}^{2s}S_+^*\;\oplus\;(\hbox{\upshape{Sym}}^{2\alpha-1}S_+^*\otimes\hbox{\upshape{Sym}}^{2\beta-1}S_-^*)$$

Consequently, we define
 $$u:\hbox{\upshape{Sym}}^{2\alpha}S_+^*\otimes\hbox{\upshape{Sym}}^{2\beta}S_-^*\longrightarrow \hbox{\upshape{Sym}}^{2\alpha-1}S_+^*\otimes\hbox{\upshape{Sym}}^{2\beta-1}S_-^*$$
proceeding as follows. First, extend $e^0\in V^*$ by $\mathbb{C}$-linearity to obtain an element $e^0\in V_{\mathbb{C}}^*$. Then compose with $\Gamma_{\mathbb{C}}:S_+^*\otimes S_-^*\longrightarrow V_{\mathbb{C}}$. This gives a map $e^0\circ\Gamma_{\mathbb{C}}:S_+^*\otimes S_-^*\longrightarrow \mathbb{C}$.\\ Now let $\iota:\hbox{\upshape{Sym}}^{2\alpha}S_+^*\otimes\hbox{\upshape{Sym}}^{2\beta}S_-^*\hookrightarrow S_+^*\otimes S_-^*\otimes  \hbox{\upshape{Sym}}^{2\alpha-1}S_+^*\otimes\hbox{\upshape{Sym}}^{2\beta-1}S_-^*$ be the canonical inclusion. Finally, set $u:=((e^0\circ\Gamma_{\mathbb{C}})\otimes\hbox{\upshape{Id}})\circ \iota$. Then $u$ is $\hbox{SU}(2)$-equivariant (since $e^0$ is $\hbox{SU}(2)$-equivariant).\\

 In fact, we have the following exact sequence of $\hbox{SU}(2)$-modules:
 $$0\longrightarrow \hbox{Sym}^{2s} S_+^*\longrightarrow \hbox{Sym}^{2\alpha}S_+^*\otimes \hbox{Sym}^{2\beta}S_-^* \longrightarrow \hbox{\upshape{Sym}}^{2\alpha-1}S_+^*\otimes\hbox{\upshape{Sym}}^{2\beta-1}S_-^*\longrightarrow 0$$
the second nontrivial map being $u$, and the first nontrivial map being $(\hbox{\upshape{Id}}\otimes \tilde{c}) \circ \iota_{\alpha,\beta}$, where $\iota_{\alpha,\beta}: \hbox{Sym}^{2s} S_+^*\hookrightarrow \hbox{Sym}^{2\alpha}S_+^*\otimes \hbox{Sym}^{2\beta}S_+^*$ is the canonical inclusion and $c:S_+^*\longrightarrow S_-^*$ is an $\hbox{SU}(2)$-equivariant isomorphism.\\

It is easy to check that the corresponding symbol $\;\zeta_u:\mathcal{O}_m\longrightarrow W^*\otimes E\;$ (where $E:=\hbox{\upshape{Sym}}^{2\alpha-1}S_+^*\otimes\hbox{\upshape{Sym}}^{2\beta-1}S_-^*$) is given by
$$\zeta_u(p)=((p\circ\Gamma_{\mathbb{C}})\otimes\hbox{\upshape{Id}}_E)\circ \iota$$
Let $\;\Xi:W\oplus (V^*\otimes W)\longrightarrow E\;$ be defined by $\;\Xi(w\;,\;p\otimes w)=\Xi^{(1)}(p\otimes w)\;$, where $\;\Xi^{(1)}:V^*\otimes W\longrightarrow E\;$ is given by
$$\Xi^{(1)}=(\hbox{tr}\otimes\hbox{\upshape{Id}}_E)\circ (\hbox{\upshape{Id}}_{V_{\mathbb{C}}^*}\otimes\Gamma_{\mathbb{C}}\otimes\hbox{\upshape{Id}}_E) \circ (j\otimes \iota)$$
where $j:V^*\hookrightarrow V_{\mathbb{C}}^*$ is the canonical inclusion.\\

Then $\;\Xi(w\;,\;p\otimes w)=\Xi^{(1)}(p\otimes w)=((\hbox{tr}\otimes\hbox{\upshape{Id}}_E)\circ (\hbox{\upshape{Id}}_{V_{\mathbb{C}}^*}\otimes\Gamma_{\mathbb{C}}\otimes\hbox{\upshape{Id}}_E))\;(p\otimes \iota(w))$\\
$=(\hbox{tr}\otimes\hbox{\upshape{Id}}_E)(p\otimes (\Gamma_{\mathbb{C}}\otimes\hbox{\upshape{Id}}_E)(\iota(w)))=((p\circ\Gamma_{\mathbb{C}})\otimes\hbox{\upshape{Id}}_E)(\iota(w))=\zeta_u(p)(w)$\\
for every $p\in\mathcal{O}_m$ and $w\in W$. Thus, $\zeta_u$ is the symbol of a first-order differential operator $D_u:\mathcal{C}^{\infty}(\mathbb{R}^4,W)\longrightarrow \mathcal{C}^{\infty}(\mathbb{R}^4,E)$. For $\phi\in\mathcal{C}^{\infty}(\mathbb{R}^4,W)$,
$$D_u\phi(x)=\Xi(\phi(x)\;,\;-id\phi(x))=\Xi^{(1)}(-id\phi(x))$$
We denote this ``divergence-type" differential operator $D_u$ by $\delta_{\alpha,\beta}$. In conclusion, we have the following proposition:\\

\begin{prop} \label{equations} Let $\mathcal{H}_{(m,s)}$ be the Wigner representation of mass $m>0$ and spin $s$ (obtained by induction from an irreducible unitary representation $F$ of the stabilizer $K$). Also, let $\mathbb{W}$ be the vector bundle on Minkowski spacetime $\mathbb{R}^4$ associated to the representation $W=\hbox{\upshape{Sym}}^{2\alpha}S_+^*\otimes\hbox{\upshape{Sym}}^{2\beta}S_-^*$ of the group $\hbox{\upshape{Spin}}(V)$. Assume that $F$ appears in the decomposition of $W$ under $K$ (which is equivalent to $\alpha+\beta\geq s$). Then $\mathcal{H}_{(m,s)}$ is selected by the condition $\;\zeta_u(p)(\widehat{\phi}(p))=0\;$ in momentum space, and by the following equations in spacetime:
$$\left\{\begin{array}{rcl} (\square+m^2)\phi & = & 0\\
\delta_{\alpha,\beta}\phi & = & 0  \end{array}\right.$$
where $\;\delta_{\alpha,\beta}:\mathcal{C}^{\infty}(\mathbb{R}^4,\hbox{\upshape{Sym}}^{2\alpha}S_+^*\otimes\hbox{\upshape{Sym}}^{2\beta}S_-^*)\longrightarrow \mathcal{C}^{\infty}(\mathbb{R}^4,\hbox{\upshape{Sym}}^{2\alpha-1}S_+^*\otimes\hbox{\upshape{Sym}}^{2\beta-1}S_-^*)\,.$\\
\end{prop}

\end{document}